\documentclass[a4paper,12pt]{article}
\usepackage[margin=2.5cm]{geometry}
\usepackage{setspace}
\usepackage{indentfirst}
\usepackage{footmisc}
\usepackage{amsmath}
\usepackage{amssymb}
\usepackage[nosort]{cite}
\usepackage[svgnames]{xcolor}
\usepackage[colorlinks=true,
            allcolors=.,
            bookmarksnumbered=true,
            pdfpagemode=UseNone,
            pdfstartview=FitH]{hyperref}

\newcommand{\dif}{\text{d}}

\numberwithin{equation}{section}
\makeatletter
\renewcommand\section{\@startsection {section}{1}{\z@}%
	{-3.5ex \@plus -1ex \@minus -.2ex}%
	{2.3ex \@plus.2ex}%
	{\normalfont\sffamily\bfseries}}
\renewcommand\subsection{\@startsection{subsection}{2}{\z@}%
	{-3.25ex\@plus -1ex \@minus -.2ex}%
	{1.5ex \@plus .2ex}%
	{\normalfont\sffamily\slshape}}
\makeatother

\begin{document}

\thispagestyle{empty}
\vbox{}
\vspace{2cm}

\begin{center}
  {\sffamily\LARGE{Parametric solutions to the Kerr separatrix
  }}\\[16mm]
  {\sffamily Tammy Ng~~and~~Edward Teo}
  \\[6mm]
    {\sffamily\slshape\selectfont
      Department of Physics, National University of Singapore, Singapore 117551, Singapore
    }\\[15mm]
\end{center}
\vspace{2cm}
	
\centerline{\sffamily\bfseries Abstract}
\bigskip
\noindent The Kerr separatrix is a boundary in parameter space that separates bound orbits from plunging orbits in the Kerr black hole space-time. Recently, Stein and Warburton found a polynomial equation for the location of the separatrix, for two different choices of inclination parameter. Following a method of Levin and Perez-Giz developed for the equatorial case, we use a correspondence between homoclinic orbits and unstable spherical orbits to derive explicit solutions to the separatrix polynomials. These solutions are parametrised in terms of the radius of the unstable spherical orbit.
\newpage

\section{Introduction}

Amongst the different types of geodesics in the Kerr black hole space-time, bound geodesics play a very important role in astrophysics. For example, in adiabatic models of extreme mass ratio inspirals (EMRIs), the motion of the smaller black hole (or neutron star) is modelled as a sequence of bound geodesics around the larger black hole \cite{Drasco:2005,Hughes:2021}. The parameters of the geodesic will change adiabatically due to self-force \cite{Barack:2018}, resulting in the emission of gravitational waves and the gradual inspiralling of the smaller black hole into the larger one. In the modelling of EMRIs, it is important to know when the smaller black hole will transition from being in an inspiralling orbit to one that plunges into the larger black hole. 

In the parameter space of all orbits, the boundary which separates bound orbits from those which plunge into the black hole is known as the {\it separatrix\/}. Recent studies of the Kerr separatrix include \cite{Rana:2019,Stein:2019,Compere:2021}. In particular, in \cite{Stein:2019}, Stein and Warburton found a polynomial equation which can be used to determine the location of the Kerr separatrix. It is expressed in terms of the semi-latus rectum, eccentricity and inclination of the orbit, as well as the black hole parameters. In particular, it is a 12th order polynomial in the semi-latus rectum $p$. Given the other parameters, this polynomial can be numerically solved for $p$, to obtain the location of the separatrix.

For the case of geodesics lying in the equatorial plane, the separatrix equation reduces to a fourth order polynomial \cite{OShaughnessy:2002}. In \cite{Levin:2008}, Levin and Perez-Giz derived an explicit solution to this polynomial equation. They did this by making use of the fact that orbits lying on the separatrix are {\it homoclinic orbits\/}. In the equatorial plane, a homoclinic orbit is one which approaches an unstable circular orbit in the asymptotic future and asymptotic past. These two orbits share the same energy and angular momentum, and Levin and Perez-Giz were then able to find the eccentricity and semi-latus rectum of the homoclinic orbit in terms of the radius of the unstable circular orbit and the black hole parameters. This provides an explicit solution to the separatrix polynomial, parametrised in terms of the radius of the unstable circular orbit.

In the non-equatorial case, the geodesic possesses a third conserved quantity, known as the Carter constant. In this case, a homoclinic orbit is one which approaches an unstable {\it spherical\/} orbit \cite{Wilkins:1972,Teo:2020} in the asymptotic future and asymptotic past. The method of Levin and Perez-Giz was generalised to this case in \cite{Rana:2019,Stein:2019}. In \cite{Rana:2019}, the eccentricity and semi-latus rectum of the homoclinic orbit were found explicitly in terms of the radius of the unstable spherical orbit, the Carter constant, and the black hole parameters. 

Instead of the Carter constant, it is sometimes advantageous to characterise a non-equatorial orbit by its inclination with respect to the equatorial plane. There are various choices of this inclination parameter in use in the literature; two of which that were used in \cite{Stein:2019} are $x$ and $\iota$, defined in (\ref{x}) and (\ref{iota}) below, respectively.

In this paper, we will use the method of Levin and Perez-Giz to find the eccentricity and semi-latus rectum of the homoclinic orbit in terms of the radius of the unstable spherical orbit, the inclination parameter $x$, and the black hole parameters. We then show that they provide a parametric solution to the separatrix polynomial in terms of $x$ that was found by Stein and Warburton \cite{Stein:2019}. We then do the same with the inclination angle parameter $\iota$, and find a parametric solution to the separatrix polynomial in terms of $\iota$ that was also found in \cite{Stein:2019}.

\section{Homoclinic orbits and the Kerr separatrix}
\label{review}

The Kerr metric in Boyer--Lindquist coordinates is given by
\begin{align}
\dif s^2&=-\Big(1-\frac{2Mr}{\Sigma}\Big)\,\dif t^2-\frac{4Mar\sin^2\theta}{\Sigma}\,\dif t\dif\phi\cr
&+\sin^2\theta\Big(r^2+a^2+\frac{2Ma^2r\sin^2\theta}{\Sigma}\Big)\,\dif\phi^2+\Sigma\Big(\frac{\dif r^2}{\Delta}\,+\dif\theta^2\Big)\,,
\end{align}
where
\begin{subequations}
\begin{align}
\Sigma&\equiv r^2+a^2\cos^2\theta\,,\\
\Delta&\equiv r^2-2Mr+a^2.
\end{align}
\end{subequations}
Here, $M$ and $a$ are the mass and angular momentum per unit mass of the black hole, respectively.

The motion of a particle in this space-time is governed by the geodesic equations. The geodesic equations for $r$ and $\theta$ are
\begin{subequations}
\begin{align}
\label{geodesic r}
\Sigma\frac{\dif r}{\dif\tau}&=\pm\sqrt{R(r)}\,,\\
\label{geodesic theta}
\Sigma\frac{\dif\theta}{\dif\tau}&=\pm\sqrt{\Theta(\theta)}\,,
\end{align}
\end{subequations}
where
\begin{subequations}
\begin{align}
\label{R}
R(r)&\equiv(E^2-\mu^2)r^4+2M\mu^2r^3-[a^2(\mu^2-E^2)+L_z^2]r^2+2M(aE-L_z)^2r-Q\Delta\,,\\
\label{Theta}
\Theta(\theta)&\equiv Q-\cos^2\theta\,\bigg[a^2(\mu^2-E^2)+\frac{L_z^2}{\sin^2\theta}\bigg]\,,
\end{align}
\end{subequations}
and $\tau$ is an affine parameter along the geodesic. In these expressions, $\mu$, $E$ and $L_z$ are the particle's mass, energy and angular momentum in the azimuthal direction, while $Q$ is the Carter constant characterising the particle's motion in the polar direction. Since we are interested in time-like particles, we set $\mu=1$ without loss of generality. There are also geodesic equations for $\phi$ and $t$, but they will not be needed here.

Because of the square roots in (\ref{geodesic r}) and (\ref{geodesic theta}), the particle is confined to move in regions where $R(r)\geq0$ and $\Theta(\theta)\geq0$. For bound orbits, the allowed range of $r$ lies between two adjacent roots of $R(r)$, which can be identified as the periastron $r_p$ and apastron $r_a$ of the orbit. On the other hand, the allowed range of $\theta$ is between the two roots of $\Theta(\theta)$: the minimum polar angle $\theta_-$ and the maximum polar angle $\theta_+=\pi-\theta_-$.

Since $R(r)$ is a quartic polynomial in $r$, it has four roots. We will be specifically interested in the case when $R(r)$ can be factorised as follows:
\begin{align}
\label{R1}
R(r)=(E^2-1)(r-r_4)(r-r_u)^2(r-r_a)\,,
\end{align}
with $r_4<r_u\leq r_a$. Note that there is a double root at $r=r_u$; this is where a spherical orbit can exist \cite{Wilkins:1972,Teo:2020}. If we assume $E^2<1$, this orbit is {\it unstable\/} (hence the subscript $u$ on the radius). 

There is, in fact, a bound orbit with $E^2<1$ that shares the same values of $E$, $L_z$ and $Q$ as the unstable spherical orbit. Its radius $r$ takes the range $r_p<r\leq r_a$, where $r_p=r_u$, the radius of the unstable spherical orbit. This orbit is known as a homoclinic orbit, and it describes a particle which approaches the spherical orbit in the asymptotic future and asymptotic past. In the parameter space of all orbits, homoclinic orbits form the boundary between bound orbits, and those which plunge into the black hole. This boundary is the so-called Kerr separatrix.

Homoclinic orbits lying in the equatorial plane of the Kerr black hole were studied by Levin and Perez-Giz \cite{Levin:2008}. In particular, they showed how one can relate the apastron $r_a$ to the periastron $r_p=r_u$, the parameters $E$ and $L_z$ of the unstable circular orbit at $r=r_u$, and the black hole parameters $M$ and $a$.\footnote{Levin and Perez-Giz actually set $M=1$, but this can be easily restored if necessary.} The parameters $E$ and $L_z$ themselves can be expressed in terms of $r_u$ and the black hole parameters.  

With an expression for $r_a$ in terms of $r_u$ at hand, it is possible to relate other orbital parameters, such as the eccentricity $e$ and semi-latus rectum $p$ of the orbit, to $r_u$. These two parameters are defined by
\begin{subequations}
\begin{align}
\label{r_p}
r_p&=\frac{Mp}{1+e}\,,\\
r_a&=\frac{Mp}{1-e}\,,
\end{align}
\end{subequations}
or, equivalently,
\begin{subequations}
\label{ep}
\begin{align}
e&=\frac{r_a-r_p}{r_a+r_p}\,,\\
p&=\frac{2r_ar_p}{M(r_a+r_p)}\,.
\end{align}
\end{subequations}
Levin and Perez-Giz then used (\ref{ep}) to express $e$ and $p$ in terms of $r_u$ and the black hole parameters. The result are explicit expressions for the eccentricity and semi-latus rectum of a homoclinic orbit. Moreover, they pointed out that their expressions satisfy an implicit relation involving $e$ and $p$ that was derived in \cite{OShaughnessy:2002} for the separatrix. This relation takes the form of a fourth order polynomial in $p$, with coefficients depending on $e$ and the black hole parameters. What Levin and Perez-Giz had achieved, was to find a parametric solution to the separatrix polynomial in terms of $r_u$.

To describe orbits that are not confined to the equatorial plane, a third orbital parameter is needed in addition to $e$ and $p$. There are various choices of this parameter in use in the literature. One of them is the minimum polar angle $\theta_-$, or equivalently, $z_-\equiv\cos\theta_-$. However, this parameter does not distinguish between prograde and retrograde orbits, and it turns out to be more advantageous to use the parameter \cite{Stein:2019}
\begin{align}
\label{x}
x={\rm sign}(L_z)\sin\theta_-\equiv\cos\theta_{\rm inc}\,.
\end{align}
$\theta_{\rm inc}$ can be understood as an angle of inclination measured from the equatorial plane (c.f.\ Fig.~1 of \cite{Stein:2019}). For prograde orbits, we have $x>0$ and $\theta_{\rm inc}=\pi/2-\theta_-$, while for retrograde orbits, we have $x<0$ and $\theta_{\rm inc}=\pi/2+\theta_-$. This parameter will be used in Sec.~\ref{sec:x}, when we generalise Levin and Perez-Giz's work to non-equatorial orbits.

Another commonly used inclination angle in the literature is $\iota$, defined by
\begin{align}
\label{iota}
\cos\iota=\frac{L_z}{\sqrt{L_z^2+Q}}\,.
\end{align}
For prograde orbits, we have $\cos\iota>0$ and $0\leq\iota<\pi/2$, while for retrograde orbits, we have $\cos\iota<0$ and $\pi/2<\iota\leq\pi$. We will use this parameter in Sec.~\ref{sec:iota}.

\section{Parametric solutions of the separatrix in terms of $x$}
\label{sec:x}

We begin by obtaining the relation between the Carter constant $Q$ and the parameter $x$ defined by (\ref{x}). Using the fact that $\Theta(\theta_-)=0$ and that $x^2=1-\cos^2\theta_-$, we have from (\ref{Theta}) that
\begin{align}
Q=(1-x^2)\bigg[a^2(1-E^2)+\frac{L_z^2}{x^2}\bigg]\,.
\end{align}
Substituting this into (\ref{R}) gives $R(r)$ in terms of the parameters $M$, $a$, $E$, $L_z$ and $x$.

Now, the conditions for a spherical orbit to exist at radius $r=r_u$ are that $R(r_u)=R'(r_u)=0$. These two conditions can be solved simultaneously for $E$ and $L_z$. After some simplification, we obtain
\begin{subequations}
\label{EL}
\begin{align}
E&=\frac{r_u^2(r_u-2M)+a^2r_u(1-x^2)+ax\sqrt{\Upsilon}}{\sqrt{\Gamma}}\,,\\
L_z&=\frac{-2Mar_u^2x^2+(r_u^2+a^2)x\sqrt{\Upsilon}}{\sqrt{\Gamma}}\,,
\end{align}
\end{subequations}
where
\begin{subequations}
\begin{align}
\label{Upsilon}
\Upsilon&\equiv Mr_u[r_u^2-a^2(1-x^2)]\,,\\
\Gamma&\equiv r_u\big[r_u^2+a^2(1-x^2)\big]\big[r_u^2(r_u-3M)+a^2(r_u+M)(1-x^2)+2ax\sqrt{\Upsilon}\big]\,.
\end{align}
\end{subequations}
The stability of this orbit is determined by the sign of $R''(r_u)$. Although we are primarily interested in unstable spherical orbits in this paper, we emphasize that the solution (\ref{EL}) is valid for both stable and unstable spherical orbits.

The expressions for $E$ and $L_z$ in (\ref{EL}) also apply to a homoclinic orbit with periastron $r_p=r_u$. To obtain an expression for the apastron $r_a$, we first need one for the root $r_4$ in (\ref{R1}). If we equate the constant coefficient of the polynomial $R(r)$ with that in (\ref{R1}), we obtain
\begin{align}
\label{r4}
r_4=\frac{a^2(1-x^2)[a^2x^2(1-E^2)+L_z^2]}{x^2(1-E^2)r_ar_u^2}\,.
\end{align}
If we then equate the linear coefficients of the two polynomials, we obtain a quadratic equation for $r_a$. Of the two roots of this quadratic equation, the relevant root is the larger one. After substituting (\ref{EL}) in and simplifying, we obtain the expression
\begin{align}
\label{r_a}
r_a=\frac{[r_u^2-a^2(1-x^2)]\Lambda^2+[r_u^2+a^2(1-x^2)]\sqrt{\Omega}}{-2r_u\Lambda^2+[r_u^2+a^2(1-x^2)]\Delta_u}\,,
\end{align}
where
\begin{subequations}
\begin{align}
\Lambda&\equiv\sqrt{M[r_u^2-a^2(1-x^2)]}-ax\sqrt{r_u}\,,\\
\label{Omega}
\Omega&\equiv\Lambda^4-a^2(1-x^2)\Delta_u^2\,,
\end{align}
\end{subequations}
and we have set $\Delta_u\equiv\Delta|_{r=r_u}$ for brevity, i.e.,
\begin{align}
\Delta_u&=r_u^2-2Mr_u+a^2.
\end{align}

The expression (\ref{r_a}) for $r_a$ can now be substituted into (\ref{ep}). Recalling that $r_p=r_u$, we obtain the following expressions
\begin{subequations}
\label{ep_x}
\begin{align}
\label{e_x}
e&=\frac{[3r_u^2-a^2(1-x^2)]\Lambda^2+[r_u^2+a^2(1-x^2)][\sqrt{\Omega}-r_u\Delta_u]}{[r_u^2+a^2(1-x^2)]({-}\Lambda^2+\sqrt{\Omega}+r_u\Delta_u)}\,,\\
\label{p_x}
p&=\frac{2r_u\big([r_u^2-a^2(1-x^2)]\Lambda^2+[r_u^2+a^2(1-x^2)]\sqrt{\Omega}\big)}{M[r_u^2+a^2(1-x^2)]({-}\Lambda^2+\sqrt{\Omega}+r_u\Delta_u)}\,.
\end{align}
\end{subequations}
This gives the eccentricity and semi-latus rectum of the homoclinic orbit in terms of $r_u$, as well as $M$, $a$ and $x$. In the equatorial limit $x\rightarrow\pm1$, they reduce to the expressions found in \cite{Levin:2008}. Moreover, it can be checked that $e$ and $p$ in (\ref{ep_x}) satisfy the relation $p=r_u(1+e)/M$, as they should by (\ref{r_p}).

In \cite{Stein:2019}, it was shown that the separatrix is determined by solutions to a 12th order polynomial in $p$. The coefficients of this polynomial depend on $M$, $a$, $e$ and $x$. We now show that (\ref{ep_x}) provides an explicit solution to this polynomial equation, parametrised in terms of $r_u$. 

Starting from the expression for $e$ given by (\ref{e_x}), we first solve for $\Omega$, and then substitute in the expression (\ref{Omega}) for $\Omega$. Expanding out the powers of $\Lambda$ on both sides of this equation, we obtain an equation which involves the square root $\sqrt{\Upsilon}$, where $\Upsilon$ is given by (\ref{Upsilon}).  We then solve this equation for $\Upsilon$, and then substitute in the expression (\ref{Upsilon}) for $\Upsilon$. The resulting equation is equivalent to the vanishing of a 12th order polynomial in $r_u$. It can be compactly written in the form
\begin{align}
\label{poly}
&\Big(\big\{[(2+e)r_u^2-(2-e)a^2y^2]^2-r_u^4+6r_u^2a^2y^2-a^4y^4\big\}\Delta_u\cr
&-2(r_u^2-a^2)(r_u^2+a^2y^2)[(1+e)r_u^2-(1-e)a^2y^2]\Big)^2-16a^2(1-y^2)(r_u^2+a^2y^2)\cr
&\times[(1+e)r_u^2-(1-e)a^2y^2][(1+e)^2(r_u^2+a^2y^2)-4ea^2y^2]r_u^2\Delta_u=0\,,
\end{align}
where we have set $y^2\equiv1-x^2$ for brevity.

The polynomial (\ref{poly}) can be converted to one in $p$, if we use the relation $r_u=Mp/(1+e)$. It can be checked that the 12th order polynomial found in Appendix A.1 of \cite{Stein:2019} is recovered up to an overall factor (and if we set $M=1$). This shows that (\ref{ep_x}) is indeed a parametric solution to this polynomial equation.

Finally, we remark that the parameter $r_u$ in (\ref{ep_x}) takes the range $r_{\rm ibso}\leq r_u\leq r_{\rm isso}$, where $r_{\rm ibso}$ is the radius of the innermost bound spherical orbit, and $r_{\rm isso}$ is the radius of the innermost stable spherical orbit. The upper bound $r_{\rm isso}$ can be obtained by setting $e=0$ in (\ref{poly}) and solving the resulting 12th order polynomial, while the lower bound $r_{\rm ibso}$ can be obtained by setting $e=1$ in (\ref{poly}) and solving the resulting 8th order polynomial.

\section{Parametric solutions of the separatrix in terms of $\iota$}
\label{sec:iota}

In this section, we repeat the arguments of Sec.~\ref{sec:x}, but using the parameter $\iota$ defined by (\ref{iota}) instead of $x$. The steps are similar, although the resulting expressions are somewhat longer.

We begin by solving (\ref{iota}) for $Q$, and substituting the resulting expression into (\ref{R}). This gives $R(r)$ in terms of the parameters $M$, $a$, $E$, $L_z$ and $\iota$. Solving the conditions for a spherical orbit at $r=r_u$, we obtain after some simplification,\footnote{This solution was previously found in \cite{Grossman:2011}, albeit in a different form.}
\begin{subequations}
\label{EL2}
\begin{align}
E&=\frac{Mr_u(3r_u^2-4Mr_u+a^2)a\cos\iota+\Delta_u\sqrt{\tilde\Upsilon}}{\sqrt{\tilde\Gamma}}\,,\\
L_z&=\frac{Mr_u\cos\iota\,\Xi}{\sqrt{\tilde\Gamma}}\,,
\end{align}
\end{subequations}
where
\begin{subequations}
\begin{align}
\label{tildeUpsilon}
\tilde\Upsilon&\equiv Mr_u\big[r_u^4+(2r_u^2-4Mr_u+a^2)a^2\sin^2\iota\big]\,,\\
\tilde\Gamma&\equiv M\Big\{\big[r_u^2(r_u^3-3Mr_u^2-2a^2M)+(2r_u^3+a^2r_u+a^2M)a^2\sin^2\iota\big]\Xi\cr
&\quad\qquad+2r_u(3r_u^2+a^2)\big[Mr_u(3r_u^2-4Mr_u+a^2)a\cos\iota+\Delta_u\sqrt{\tilde\Upsilon}\big]a\cos\iota\Big\}\,,\\
\Xi&\equiv r_u^4+2a^2r_u^2-4a^2Mr_u+a^4.
\end{align}
\end{subequations}

The expressions for $E$ and $L_z$ in (\ref{EL2}) also apply to a homoclinic orbit with periastron $r_p=r_u$. If we equate the constant coefficient of the polynomial $R(r)$ with that in (\ref{R1}), we obtain
\begin{align}
r_4=\frac{L_z^2a^2\tan^2\iota}{(1-E^2)r_ar_u^2}\,.
\end{align}
If we then equate the linear coefficients of the two polynomials, we obtain a quadratic equation for the apastron $r_a$. Selecting the correct root, substituting (\ref{EL2}) in and simplifying, we obtain the expression
\begin{align}
\label{r_a2}
r_a=\frac{(r_u^2-a^2)\tilde\Lambda^2+\sqrt{\tilde\Omega}}{-4r_u\tilde\Lambda^2+(r_u^2-a^2\sin^2\iota)\Xi^2}\,,
\end{align}
where
\begin{subequations}
\begin{align}
\tilde\Lambda&\equiv(r_u^2-a^2)\sqrt{M\big[r_u^4+(2r_u^2-4Mr_u+a^2)a^2\sin^2\iota\big]}-r_u^{3/2}\Delta_ua\cos\iota\,,\\
\label{tildeOmega}
\tilde\Omega&\equiv(r_u^2-a^2)\big\{(r_u^2-a^2)\tilde\Lambda^4+\big[4r_u\tilde\Lambda^2-(r_u^2-a^2\sin^2\iota)\Xi^2\big]a^2\sin^2\iota\,\Xi^2\big\}\,.
\end{align}
\end{subequations}

The expression (\ref{r_a2}) for $r_a$ can now be substituted into (\ref{ep}). With $r_p=r_u$, we obtain%
\begin{subequations}
\label{ep_iota}
\begin{align}
\label{e_iota}
e&=\frac{(5r_u^2-a^2)\tilde\Lambda^2+\sqrt{\tilde\Omega}-r_u(r_u^2-a^2\sin^2\iota)\Xi^2}{-(3r_u^2+a^2)\tilde\Lambda^2+\sqrt{\tilde\Omega}+r_u(r_u^2-a^2\sin^2\iota)\Xi^2}\,,\\
\label{p_iota}
p&=\frac{2r_u}{M}\frac{(r_u^2-a^2)\tilde\Lambda^2+\sqrt{\tilde\Omega}}{-(3r_u^2+a^2)\tilde\Lambda^2+\sqrt{\tilde\Omega}+r_u(r_u^2-a^2\sin^2\iota)\Xi^2}\,.
\end{align}
\end{subequations}
This gives the eccentricity and semi-latus rectum of the homoclinic orbit in terms of $r_u$, as well as $M$, $a$ and $\iota$. 

In \cite{Stein:2019}, the parameter $\iota$ was also used to derive a 12th order separatrix polynomial in $p$. We now show that (\ref{ep_iota}) provides an explicit solution to this polynomial equation, parametrised in terms of $r_u$. 

Starting from the expression for $e$ given by (\ref{e_iota}), we first solve for $\tilde\Omega$, and then substitute in the expression (\ref{tildeOmega}) for $\tilde\Omega$. Expanding out the powers of $\tilde\Lambda$ on both sides of this equation, we obtain an equation which involves the square root $\sqrt{\tilde\Upsilon}$, where $\tilde\Upsilon$ is given by (\ref{tildeUpsilon}).  We then solve this equation for $\tilde\Upsilon$, and then substitute in the expression (\ref{tildeUpsilon}) for $\tilde\Upsilon$. The resulting equation is equivalent to the vanishing of a 12th order polynomial in $r_u$. It can be compactly written in the form
\begin{align}
\label{poly2}
&\Big([(1+e)(3+e)r_u^4-(1-e)(3-e)a^4\sin^2\iota]\Delta_u\cr
&+2(r_u^2-a^2)\big\{(1+e)r_u^4+[2er_u^2-(1-e)a^2]a^2\sin^2\iota\big\}\Big)^2\cr
&-4(1+e)[(3+e)r_u^2-(1-e)a^2]\Big((1-e)[(1+e)r_u^2-(3-e)a^2\sin^2\iota]a^2\sin^2\iota\,\Delta_u\cr
&+2(r_u^2-a^2\sin^2\iota)\big\{(1+e)r_u^4+[2er_u^2-(1-e)a^2]a^2\sin^2\iota\big\}\Big)r_u^2\Delta_u=0\,.
\end{align}

The polynomial (\ref{poly2}) can be converted to one in $p$, if we use the relation $r_u=Mp/(1+e)$. It can be checked that the 12th order polynomial found in Appendix A.2 of \cite{Stein:2019} is recovered up to an overall factor (and if we set $M=1$). This shows that (\ref{ep_iota}) is indeed a parametric solution to this polynomial equation.

\section{Conclusion}

In this paper, we have used a correspondence between homoclinic orbits and unstable spherical orbits to obtain parametric solutions to the 12th order separatrix polynomials found by Stein and Warburton \cite{Stein:2019}. The solution (\ref{ep_x}) is expressed in terms of the orbital inclination parameter $x$, while the solution (\ref{ep_iota}) is expressed in terms of the inclination angle $\iota$. The solutions also depend on the radius of the unstable spherical orbit $r_u$, or equivalently, the periastron of the homoclinic orbit.

In practice, such as in the modelling of EMRIs, the separatrix polynomial is used to solve for $p$ numerically, given the parameters $M$, $a$, $e$ and $x$ (or $\iota$). This has been implemented in the Black Hole Perturbation Toolkit \cite{BHPT} by Stein and Warburton \cite{Stein:2019}. In fact, it replaces an older, less efficient implementation based on a generalisation of Levin and Perez-Giz's work \cite{Levin:2008}. Since the approach in this paper is also based on a generalisation of \cite{Levin:2008}, it is unlikely to provide a faster way to solve for $p$ numerically given the other parameters. However, it is worth exploring if the exact solution we have found could open up new---and possibly more efficient---ways to check if the inspiralling black hole has crossed the separatrix.

In any case, there are clearly benefits to having explicit solutions to the separatrix polynomial at hand. For example, they provide a simple and efficient way to plot out the separatrix surface, without having to deal with unphysical solutions that may arise when the 12th order polynomial is solved directly \cite{Stein:2019}. More importantly, they open up the possibility of studying the properties of the separatrix surface analytically. We leave these interesting questions for future work.

\section*{Acknowledgements}

ET wishes to thank Alvin Chua, Soichiro Isoyama and especially Josh Mathews for useful discussions.


\bigskip\bigskip


\begin{thebibliography}{99}
\setlength{\baselineskip}{19pt}

\bibitem{Drasco:2005}
S.~Drasco and S.~A.~Hughes,
``Gravitational wave snapshots of generic extreme mass ratio inspirals,''
Phys. Rev. D \textbf{73} (2006) 024027
[\href{https://doi.org/10.1103/PhysRevD.73.024027}{\tt doi:10.1103/PhysRevD.73.024027}; \href{http://arxiv.org/abs/gr-qc/0509101}{\tt
arXiv:gr-qc/0509101}].

\bibitem{Hughes:2021}
S.~A.~Hughes, N.~Warburton, G.~Khanna, A.~J.~K.~Chua and M.~L.~Katz,
``Adiabatic waveforms for extreme mass-ratio inspirals via multivoice decomposition in time and frequency,''
Phys. Rev. D \textbf{103} (2021) 104014
[\href{https://doi.org/10.1103/PhysRevD.103.104014}{\tt doi:10.1103/PhysRevD.103.104014}; \href{http://arxiv.org/abs/2102.02713}{\tt
arXiv:2102.02713 [gr-qc]}].

\bibitem{Barack:2018}
L.~Barack and A.~Pound,
``Self-force and radiation reaction in general relativity,''
Rept. Prog. Phys. \textbf{82} (2019) 016904
[\href{https://doi.org/10.1088/1361-6633/aae552}{\tt doi:10.1088/1361-6633/aae552}; \href{http://arxiv.org/abs/1805.10385}{\tt arXiv:1805.10385 [gr-qc]}].

\bibitem{Rana:2019}
P.~Rana and A.~Mangalam,
``Astrophysically relevant bound trajectories around a Kerr black hole,''
Class.\ Quant.\ Grav.\  {\bf 36} (2019) 045009
[\href{https://doi.org/10.1088/1361-6382/ab004c}{\tt doi:10.1088/1361-6382/ab004c}; \href{http://arxiv.org/abs/1901.02730}{\tt arXiv:1901.02730 [gr-qc]}].

\bibitem{Stein:2019}
L.~C.~Stein and N.~Warburton,
``Location of the last stable orbit in Kerr spacetime,''
Phys.\ Rev.\ D \textbf{101} (2020) 064007
[\href{https://doi.org/10.1103/PhysRevD.101.064007}{\tt doi:10.1103/PhysRevD.101.064007}; \href{http://arxiv.org/abs/1912.07609}{\tt arXiv:1912.07609 [gr-qc]}].

\bibitem{Compere:2021}
G.~Comp\`ere, Y.~Liu and J.~Long,
``Classification of radial Kerr geodesic motion,''
Phys. Rev. D \textbf{105} (2022) 024075
[\href{https://doi.org/10.1103/PhysRevD.105.024075}{\tt doi:10.1103/PhysRevD.105.024075}; \href{http://arxiv.org/abs/2106.03141}{\tt
[arXiv:2106.03141 [gr-qc]}].

\bibitem{OShaughnessy:2002}
R.~O'Shaughnessy,
``Transition from inspiral to plunge for eccentric equatorial Kerr orbits,''
Phys. Rev. D \textbf{67} (2003) 044004
[\href{https://doi.org/10.1103/PhysRevD.67.044004}{\tt doi:10.1103/PhysRevD.67.044004}; \href{http://arxiv.org/abs/gr-qc/0211023}{\tt arXiv:gr-qc/0211023}].

\bibitem{Levin:2008}
J.~Levin and G.~Perez-Giz,
``Homoclinic orbits around spinning black holes. I. Exact solution for the Kerr separatrix,''
Phys.\ Rev.\ D \textbf{79} (2009) 124013
[\href{https://doi.org/10.1103/PhysRevD.79.124013}{\tt doi:10.1103/PhysRevD.79.124013}; \href{http://arxiv.org/abs/0811.3814}{\tt arXiv:0811.3814 [gr-qc]}].

\bibitem{Wilkins:1972}
D.~C.~Wilkins,
``Bound geodesics in the Kerr metric,''
Phys.\ Rev.\ D {\bf 5} (1972) 814
[\href{https://doi.org/10.1103/PhysRevD.5.814}{\tt doi:10.1103/PhysRevD.5.814}].

\bibitem{Teo:2020}
E.~Teo,
``Spherical orbits around a Kerr black hole,''
Gen. Rel. Grav. \textbf{53} (2021) 10
[\href{https://doi.org/10.1007/s10714-020-02782-z}{\tt doi:10.1007/s10714-020-02782-z}; \href{http://arxiv.org/abs/2007.04022}{\tt arXiv:2007.04022 [gr-qc]}].

\bibitem{Grossman:2011}
R.~Grossman, J.~Levin and G.~Perez-Giz,
``The harmonic structure of generic Kerr orbits,''
Phys. Rev. D \textbf{85} (2012) 023012
[\href{https://doi.org/10.1103/PhysRevD.85.023012}{\tt doi:10.1103/PhysRevD.85.023012}; \href{http://arxiv.org/abs/1105.5811}{\tt arXiv:1105.5811 [gr-qc]}].

\bibitem{BHPT}
Black Hole Perturbation Toolkit [\href{https://bhptoolkit.org}{\tt bhptoolkit.org}].

\end{thebibliography}
\end{document}